\documentstyle[aps,twocolumn,epsfig]{revtex}
\begin{document}
\newcommand{\ve}[1]{\mbox{\boldmath $#1$}}
\twocolumn[\hsize\textwidth\columnwidth\hsize
\csname@twocolumnfalse%
\endcsname

\draft
 
\title {Solitary waves in clouds of Bose-Einstein condensed atoms}
\author{A. D. Jackson,$^1$ G. M. Kavoulakis,$^2$ and C. J. Pethick$^{2,3}$}
\date{\today} 
\address{$^1$Niels Bohr Institute, Blegdamsvej 17, DK-2100 Copenhagen \O,
        Denmark, \\
        $^2$Nordita, Blegdamsvej 17, DK-2100 Copenhagen \O, Denmark, \\
        $^3$Department of Physics, University of Illinois at
        Urbana-Champaign, 1110 West Green Street, Urbana, Illinois 61801-3080} 
\maketitle
 
\begin{abstract}

       We consider the conditions under which solitary waves can exist
in elongated clouds of Bose-Einstein condensed atoms. General expressions
are derived for the velocity, characteristic size, and spatial profile
of solitary waves, and the low- and high-density limits are examined. 

\end{abstract}
 
\vskip2pc]

Clouds of Bose-Einstein condensed atoms in elongated traps provide 
excellent conditions for investigating the propagation of essentially 
one-dimensional sound pulses under \cite{andrews}.  In previous work, 
propagation of pulses in such traps was considered in the Thomas-Fermi 
approximation, and it was demonstrated that pulses propagate at a speed 
of $(\bar{n} U_0/m)^{1/2}$ in the linear regime.  Here, $U_0=4\pi \hbar^2 
a_{\rm sc}/m$ is the effective two-body interaction matrix element,
$\bar{n}$ is the particle density averaged over the cross section of
the cloud \cite{zaremba,KP}, $m$ is the atomic mass, and $a_{\rm sc}$ is 
the scattering length for atom-atom collisions.  Non-linear effects were 
also investigated in Ref.\,\cite{KP} and were found to be important for 
conditions of experimental relevance.  The effects of dispersion were 
neglected in this study since the length scales of interest were much 
larger than the superfluid coherence length, which sets the scale on which 
dispersive effects become important.  It is of interest to include the
effects of dispersion and to investigate the possible existence of solitary
waves in these systems. 

Solitary waves have been studied in many different physical contexts 
\cite{chu,rajaraman}.  We shall not attempt to review the extensive 
literature for the Gross-Pitaevskii equation, the non-linear Schr\"odinger 
equation that describes motion of a Bose-Einstein condensate, apart from 
mentioning the pioneering work of Zakharov and Shabat \cite{Rus} on solitary 
waves in bulk systems.  There are a number of studies for clouds of 
Bose-Einstein condensed atoms in a trap.  Morgan {\it et al.} \cite{morgan} 
considered a class of solitary wave solutions which are, in effect, 
center-of-mass oscillations of the stationary states of the time-independent 
Gross-Pitaevskii equation.  
 
It is possible experimentally to make very elongated clouds of Bose-Einstein 
condensed atoms in magnetic or optical traps, and in this paper we examine the 
possibility of solitary waves propagating along the axis of such a 
cloud. The solutions we seek are the analogs of the solitary waves first 
observed by Scott Russell on his historic horse ride along the canal
\cite{srussell}.  However, 
an important difference between canals and Bose-Einstein condensed clouds is 
that, whereas it is often a good approximation to assume that the depth of 
water in a canal is constant and thus the sound speed is independent of 
position, this
is a poor approximation in Bose-Einstein condensed clouds, where the density, 
and hence also the sound speed, vary significantly in directions transverse to 
the axis of the trap.  Consequently, the conditions under which one can find 
solitary wave solutions for Bose-Einstein condensed clouds is more restricted 
than for canals.
We wish to consider in this paper situations where the variation of cloud 
properties along the axis of the trap may be neglected.  In the directions 
perpendicular to the axis, we shall assume that the cloud is confined by a 
harmonic trapping potential.  However, to set the stage and establish notation, 
let us first consider the case of a homogeneous medium and
then generalize our discussion to atoms in traps which confine particles in 
two directions.
 
The condensate wavefunction $\Psi$ satisfies the Gross-Pitaevskii equation, 
\begin{eqnarray}
   i \hbar \frac {\partial \Psi} {\partial t} = 
 \left( -\frac {\hbar^2 {\ve \nabla}^2} {2m} + U_0 |\Psi|^2 + V \right) \Psi,
\label{GPequ}
\end{eqnarray}
where $V$ is the external potential, which may be taken to be zero for
a homogeneous medium.  We write the condensate wavefunction as
$\Psi = n^{1/2} e^{i \phi}$, where $n$ is the particle density, which 
is spatially uniform in equilibrium and in the absence of an external 
potential.  To obtain equations resembling those of hydrodynamics, it is
convenient to work in terms of a superfluid velocity, ${\bf v} = \hbar
{\ve \nabla} \phi/m$.  We consider solutions that depend on a single
coordinate, $z$.  From the real and imaginary parts of the Gross-Pitaevskii 
equation one obtains two equations, the equation of
continuity 
\begin{eqnarray}
   \frac {\partial n} {\partial t} = - \frac {\partial (n v)}
  {\partial z} \ ,
\label{cont1d}
\end{eqnarray}
and the equation for the phase,
\begin{eqnarray}
   \hbar \frac {\partial \phi} {\partial t} = 
   -n U_0 - \frac 1 2 m v^{2} + 
 \frac {\hbar^2} {2 m n^{1/2}} \frac {\partial^{2} n^{1/2}} {\partial z^2} \ .
\label{phiequ}
\end{eqnarray} 
When differentiated with respect to $z$, the latter gives the generalized
Euler equation, 
\begin{eqnarray}
   m  \frac {\partial v} {\partial t} = - \frac {\partial} {\partial z}
  \left( \mu(n) + \frac 1 2 m v^2  - \frac {\hbar^2} {2m n^{1/2}}
 \frac {\partial^{2} n^{1/2}} {\partial z^2} \right) \ ,
\label{nv1d}
\end{eqnarray}
where we have introduced the chemical potential, $\mu$.  For a dilute 
Bose gas, $\mu(n) = n U_0$.  We seek solutions to these equations for which
the fluid velocity and particle density propagate at a uniform velocity, $u$,
without change of form, i.e., they depend on $z$ and $t$ only through
the combination $z-ut$. 
Thus, one may write $\partial v/ \partial t = -u (\partial v/ \partial z)$ 
and $\partial n/ \partial t = -u (\partial n/ \partial z)$.  Far away from 
the solitary wave, the condensate is at rest and has its equilibrium density, 
$n_0$.  With these boundary conditions, the continuity equation, 
Eq.\,(\ref{cont1d}), gives
\begin{eqnarray}
    v= u \left( 1 - \frac {n_0} {n} \right) \ .
\label{eq1}
\end{eqnarray}
Also, Eq.\,(\ref{nv1d}) can be written as
\begin{eqnarray}
   -m  \frac {\partial v} {\partial z} = - \frac {\partial} {\partial z}
  \left( \mu(n) + \frac 1 2 m v^2  - \frac {\hbar^2} {2m n^{1/2}}
 \frac {\partial^{2} n^{1/2}} {\partial z^2} \right) \ .
\label{nv1dint}
\end{eqnarray}
We integrate this equation to obtain
\begin{eqnarray}
    \frac {\hbar^2} {2m n^{1/2}}
  \frac {\partial^2 n^{1/2}} {\partial z^2} =
 (n - n_0) U_0 +
 \frac m 2  (v-u)^2 -
\frac m 2  u^2 \ .
\label{eq2}
\end{eqnarray}
Here, we have added the integration constant $-n_0 U_0$ to impose the 
boundary condition $n \rightarrow n_0$ at infinity.  Combining 
Eqs.\,(\ref{eq1}) and (\ref{eq2}), we obtain a differential
equation for $n$. If we multiply this equation by $\partial n^{1/2}/
\partial z$ and integrate with respect to $z$, we find
\begin{eqnarray}
     \frac {\hbar^2} {2m}
   \left( \frac {\partial n^{1/2}} {\partial z} \right)^2 =
  (n U_0 - m u^2) \frac {(n - n_0)^2} {2n} \ .
\label{poles1d}
\end{eqnarray}
We note that the phase has the more general form $\phi(z,t) = 
\phi_1(z-vt)+\phi_2(t)$, which can be seen from the equation $\partial \phi/
\partial z =mv/\hbar$.  Thus, Eq.\,(\ref{phiequ}), can be written as
\begin{eqnarray}
   m \frac {\partial \phi_2} {\partial t} =
  \frac {\hbar^2} {2 m n^{1/2}} \frac {\partial^2 n^{1/2}} {\partial z^2}
 - n U_0 + \frac m 2 (v-u)^2 - \frac m 2 u^2 \ .
\label{phiequ2}
\end{eqnarray}
Since the left side of this equation is a function of $t$ and the right
is a function of $z-ut$, each must be equal to a constant.  This constant 
must equal $-n_0 U_0$ in order to satisfy the boundary condition $n 
\rightarrow n_0$ at infinity.  Thus, $\phi_2(t)=-n_0 U_0 t$ is a linear 
function of time.

We see from Eq.\,(\ref{poles1d}) that the condition $n U_0 - mu^2 \ge 0$ 
must be satisfied in order to obtain real solutions.  In other words, 
the density must exceed the minimum value $n_{\rm{min}}$,
\begin{eqnarray}
   n_{\rm{min}} = \frac {m u^2} {U_0} \ .
\label{nmin}
\end{eqnarray}
To obtain solutions that are localized in space, $n$ must lie between
$n_{\rm{min}}$ and $n_0$. This has a ready interpretation in terms of 
one-dimensional motion of a classical particle whose spatial coordinate
is proportional to $n^{1/2}$ if $z$ is regarded as the time variable.
The classical potential is then proportional to $-(n-n_{\rm{min}})(n-n_0)^2$,
and the solitary wave solutions correspond to an oscillation of the 
``particle'' from $n=n_0$ to $n=n_{\rm{min}}$ and back again. Thus,
solitary waves for this problem are depressions in the density.  In
contrast, solitary waves in canals correspond to elevations of the surface 
of the water. We also note that the velocity of the wave is equal to the 
sound speed at the minimum density, $n_{\rm{min}}$. The sound speed in a 
uniform gas with density $n_0$ is given by $c^2 = n_0 U_0/m$.
It follows from Eq.\,(\ref{nmin}) that 
\begin{eqnarray}
     \frac {n_{\rm{min}}} {n_0} = \frac {u^2} {c^2}\ .
\label{ratio}
\end{eqnarray}

Integrating Eq.\,(\ref{poles1d}) we find that the profile of the wave is 
given by
\begin{eqnarray}
    n(z) = n_{\rm{min}} + (n_0 - n_{\rm{min}}) \tanh^2(z/\zeta)\ ,
\label{profile1d}
\end{eqnarray}
where $\zeta = 2^{1/2} \xi(n_0) [1-(n_{\rm min}/ n_0)]^{-1/2}$ and 
$\xi(n_0)$ is the coherence length corresponding to the background
density $n_0$, $\xi(n_0)=(8 \pi n_0 a_{\rm{sc}})^{-1/2}$.  Therefore, 
$\zeta$, which gives the spatial extent of the solitary wave, is on the 
order of the coherence length that corresponds to the background density.  
Figure 1 shows the profile of two solitary waves, $n(z)$, for Na atoms with a 
background density of $n_0 = 10^{13}$ cm$^{-3}$, and for two
values of $n_{\rm{min}}/n_0$, 20\% (solid curve) and 80\% (dashed curve).
\begin{figure}
\begin{center}
\epsfig{file=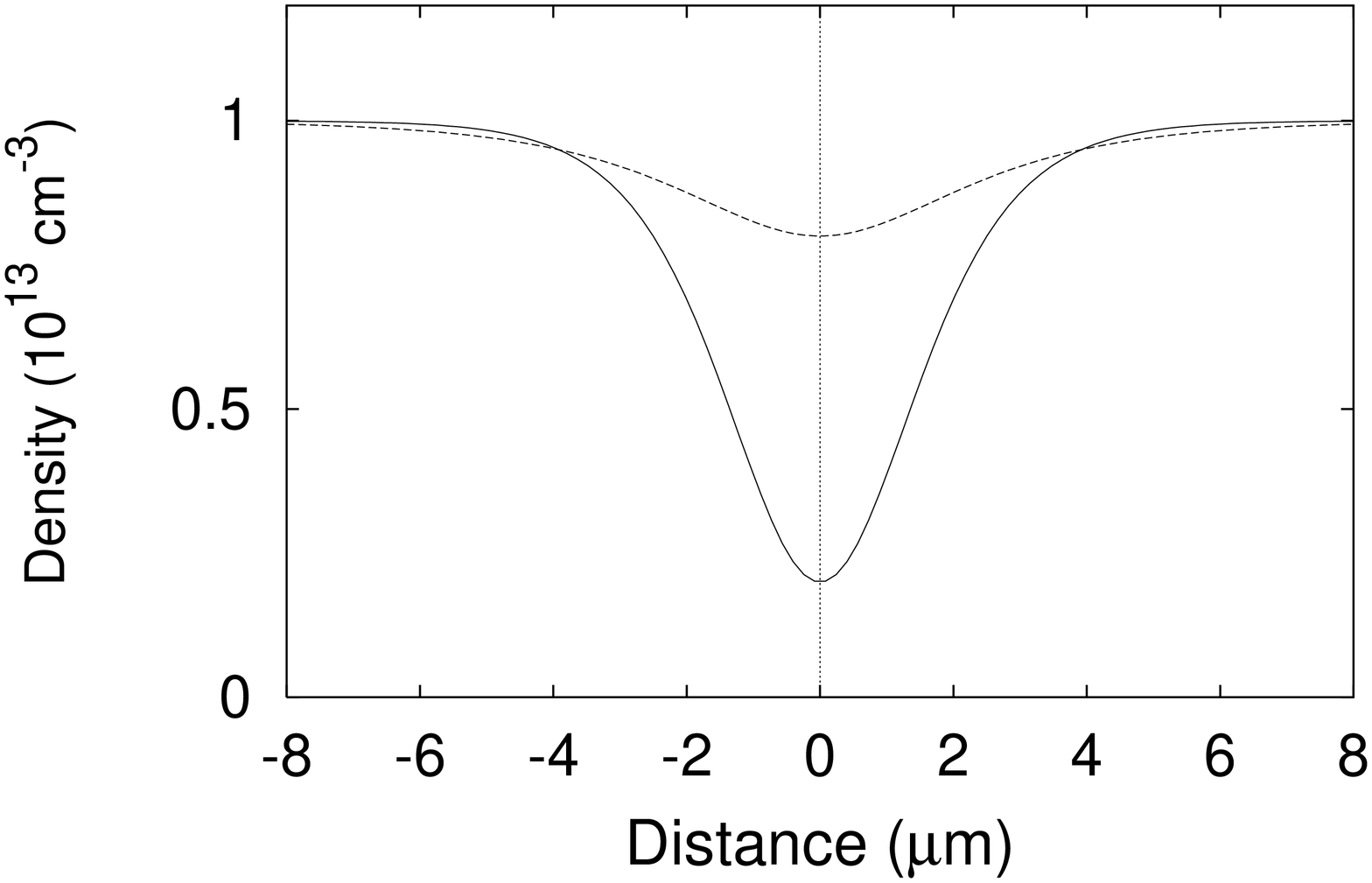,width=8cm,height=8cm,angle=-90}
\begin{caption}
{Density profiles of solitary waves as function of space, as given by
Eq.\,(\ref{profile1d}).  Here the background density of atoms is $n_0
= 10^{13}$ cm$^{-3}$ and therefore for Na atoms $\xi(n_0) \approx 1.2$ $\mu$m.
For the solid curve, the ratio $n_{\rm min}/n_0$ = 20\% and $\zeta \approx
1.9$ $\mu$m; for the dashed curve, $n_{\rm min}/n_0$ = 80\% and $\zeta
\approx 3.8$ $\mu$m.}
\end{caption}
\end{center}
\label{FIG1}
\end{figure}
\noindent
For typical densities in the MIT experiment \cite{andrews} ($n \approx 
10^{14}$ cm$^{-3}$ at the center of the cloud), $\xi$ is $\approx 0.2 - 0.4$ 
$\mu$m with $a_{\rm sc} = 27.5$ \AA \, for Na atoms \cite{Dav}. However,
in order to be able to observe solitary waves it is desirable to look
at lower densities than these, so that the coherence length becomes sufficiently
large that structures can be resolved by optical means. For non-zero 
$n_{\rm min}$ the solutions (\ref{profile1d}) 
correspond to what are termed ``gray solitons'' 
in Ref.\,\cite{Scott}.  Finally, we point out that, if $u=0$ ($n_{\rm min} 
= 0$), Eq.\,(\ref{profile1d}) gives the well-known kink solution 
$\Psi = n_0^{1/2} \tanh(z/\zeta)$ \cite{Pitaevskii}, which is sometimes 
referred to as a ``dark soliton'' \cite{Scott,Dum}.

We turn now to the more realistic problem of atoms in a trapping potential 
which is harmonic in the transverse direction and for which there is no 
restoring force along the axis of the trap, which we take to be the $z$ axis.
We assume that the transverse dimension of the cloud is so small
that the time scale for adjustment of the transverse profile of the
particle density to the equilibrium form appropriate for the instantaneous 
number of particles per unit length is small compared with the time for the 
pulse to pass a given point. Later, we shall investigate what this condition 
means quantitatively. The problem becomes one-dimensional, and the solitary 
pulse may be specified in terms of a local velocity, $v(z)$, and a local 
density of particles per unit length, $\sigma(z)$,
\begin{eqnarray}
     \sigma(z)=\int dx dy \, |\Psi(x,y,z)|^2 \ .
\label{sigma}
\end{eqnarray}
Here, $x$ and $y$ are coordinates perpendicular to the axis of the trap.
With this assumption, the wavefunction may be written in the form
\begin{eqnarray}
   \Psi({\ve r},t)=f(z,t) \, g(x,y,\sigma)\ ,
\label{wavefunction}
\end{eqnarray}
where $g$ is the equilibrium wavefunction for the transverse motion;
$g(x,y,\sigma)$ depends on time implicitly through the time
dependence of $\sigma$. We choose $g$ to be normalized so that 
$\int |g|^2 dx dy = 1$, and therefore from Eqs.\,(\ref{sigma})
and (\ref{wavefunction}), $|f|^2=\sigma$.

It is convenient to derive the equations for $f$ and $g$ from a
variational principle. The Ginzburg-Pitaevskii equation may be
derived by requiring stationarity of the action
\begin{eqnarray}
    {\cal{S}} = -\int
   \frac {i \hbar} 2 \left( \Psi^* \frac {\partial \Psi} {\partial t} -
 \Psi \frac {\partial \Psi^*} {\partial t} \right)  d{\ve r} dt+ 
\phantom{XXXXX}
 \nonumber \\
+\int \left( \frac {\hbar^2} {2m} |{\ve \nabla} \Psi|^2 +
  \frac 1 2 U_0 |\Psi|^4 + V |\Psi|^2 \right) d{\ve r} dt \ .
\label{energy}
\end{eqnarray}
Using Eq.\,(\ref{wavefunction}) for $\Psi$, we can write  
Eq.\,(\ref{energy}) as
\begin{eqnarray}
    {\cal{S}} = 
   -\int \frac {i \hbar} 2 \left( (fg)^* \frac {\partial (fg)} {\partial t} -
  fg \frac {\partial (fg)^*} {\partial t} \right)  d{\ve r} dt+ \nonumber \\
  + \int \left[ \frac {\hbar^2} {2m} 
   \left( \left| g \frac {\partial f} {\partial z} \right|^2 + 
 |f {\ve \nabla}_{\perp} g|^2 \right) \right]  d{\ve r} dt
    +  \phantom{X} \nonumber \\  
  + \int \left[ \frac 1 2 U_0 |fg|^4 + V |fg|^2 \right] d{\ve r} dt \ ,  
\label{energy2}
\end{eqnarray}
where ${\ve \nabla}_{\perp}^2 = \partial^2/\partial x^2 + 
\partial^2/\partial y^2$.
Terms containing $(\partial f/\partial z)(\partial g/\partial z)$
vanish because of the normalization condition on $g$. Minimizing
Eq.\,(\ref{energy2}) with respect to $g^*$, we find that $g$ obeys 
the equation 
\begin{eqnarray}
  - \frac {\hbar^2 {\ve \nabla}_{\perp}^2} {2m} g + V g 
+U_0 |f g|^2 g+ \phantom{XXXXXXXXX} \nonumber \\ \phantom{XXXXXXXXX} 
 + \frac {\hbar^2} {2m|f|^2} \left|
 \frac {\partial f} {\partial z} \right|^2 g =  \mu(\sigma) g \ ,
\label{equilequf}
\end{eqnarray}
where $\mu$ is the chemical potential.  We neglect the last term on 
the left side of Eq.\,(\ref{equilequf}) since, as we show below, the
characteristic length of pulses is sufficiently long that it is
negligible in all cases of interest.  We now minimize Eq.\,(\ref{energy2}) 
with respect to $f^*$ and find
\begin{eqnarray}
 i \hbar \frac {\partial f} {\partial t} =
  - \frac {\hbar^2} {2m} \frac {\partial^2 f} {\partial z^2} + 
 \left( \frac {\hbar^2} {2m} \int |{\ve \nabla}_{\perp} g|^2 \, dx dy \right) f
+ \nonumber \\  + U_0 \left( \int |g|^4 dx dy
  \right) |f|^2 f + \left( \int |g|^2 V dx dy \right)f \ .
\label{gp1}
\end{eqnarray}
Following the same procedure as for the homogeneous medium, we assume that 
$f=\sigma^{1/2} e^{i \phi}$, with the velocity field associated with $f$ 
being $v=(\hbar/m) \partial \phi/ \partial z$.  Again, we obtain hydrodynamic 
equations for $\sigma$ and $v$, which are the same as Eqs.\,(\ref{cont1d}) 
and (\ref{nv1d}) but with $n$ replaced by $\sigma$ and $\mu$ given in the 
present case by
\begin{eqnarray}
 \mu(\sigma)&=&  \left( \frac {\hbar^2} {2m} 
    \int |{\ve \nabla}_{\perp} g|^2 \, dx dy \right) 
+ \nonumber \\  &+& \left( \int |g|^2 V dx dy \right) +
 U_0 \left( \int |g|^4 dx dy \right) \sigma \ .
\label{mu}
\end{eqnarray}
The first term on the right side of Eq.\,(\ref{mu}) is the kinetic 
energy in the transverse direction, the second is the potential energy due
to the confining potential, and the third is the energy due to 
interactions between atoms.  Assuming that $\sigma$ and $v$ are 
functions of $z-ut$ only with $u$ constant, we find that
\begin{eqnarray}
     \frac {\hbar^2} {2m} 
   \left( \frac {\partial \sigma^{1/2}} {\partial z} \right)^2 =
  [\epsilon(\sigma) - \epsilon(\sigma_0)] - \nonumber \phantom{XXXXXXX} \\
  \mu(\sigma_0) (\sigma - \sigma_0)
 -  m u^2 \frac {(\sigma - \sigma_0)^2} {2\sigma} \ ,
\label{poles}
\end{eqnarray}
where $\epsilon(\sigma) = \int_0^{\sigma} \mu(\sigma') d\sigma'$
is the energy per unit length. In Eq.\,(\ref{poles}), we have imposed 
the boundary conditions $\sigma=\sigma_0$ and $v=0$ far away from the 
disturbance.

There is no simple expression for the energy per unit length for general 
values of $\sigma$.  To obtain analytical results, we explore some limiting 
cases for the experimentally relevant situation where the confining potential
in the transverse directions is harmonic and rotationally invariant, and 
given by $V=m \omega_{\perp}^2 (x^2+y^2)/2$.
In the low-density regime, the interaction energy can be treated 
perturbatively and the problem reduces essentially to the one-dimensional 
case treated above.  As we shall show below, in this limit $\epsilon(\sigma) 
\propto \sigma (1 + a_{\rm sc} \sigma)$.  In the high-density limit, the 
Thomas-Fermi approximation applies \cite{bp}, and $\epsilon(\sigma)$
varies as $\sigma^{3/2}$.

Before examining the two cases separately, we estimate the density, 
$\sigma_c$, at which the cross-over between the two limits occurs.
This corresponds to the condition that the interaction energy per particle 
is on the order of the oscillator energy in the transverse direction.  
The interaction energy per particle is $n U_0$.  If $A$ is the cross section 
of the cloud and $R_{\perp}$ is the corresponding radius, $n = \sigma /A$.  
Therefore, the condition determining $\sigma_c$ is, $U_0 \sigma_c /A \sim 
\hbar \omega_{\perp}$. Denoting the characteristic length scale for the ground
state of a particle in the transverse confining potential by $a_{\perp} = 
(\hbar/m\omega_{\perp})^{1/2}$ and assuming that $R_{\perp}=a_{\perp}$,
the cross section of the cloud is $A = \pi a_{\perp}^2$; thus one 
finds $\sigma_c \sim a_{\rm sc}^{-1}$, which gives $\sigma_c \approx 
4 \times 10^6$ cm$^{-1}$ for Na atoms.  Alternatively, we can determine 
$\sigma_c$ by equating the interaction energy per particle and the kinetic 
energy of the atoms due to their confinement in the transverse direction, 
$\sigma_c U_0/A \sim \hbar^2/(mA)$, which gives the same result for 
$\sigma_c$.  This expression for $\sigma_c$ is equivalent to the condition 
that the coherence length, $\xi$, be comparable to the transverse dimension 
of the cloud, $ R_{\perp}$. 

We now examine the low-density limit, $\sigma \ll \sigma_c$ in which 
the interaction energy can be treated perturbatively.  To find the 
differential equation that $g$ satisfies in this limit, we neglect the 
interaction-energy term (the third term on the left side) in 
Eq.\,(\ref{equilequf}).  The last term on the left side is
$\sim \hbar^2/(2 m \zeta^2) (\delta f/f)^2$, where $\delta f$ is the 
deviation of $f$, which is much less than $\hbar^2/(2 m a_{\perp}^2)$ because
of the condition $\zeta \gg a_{\perp}$. Therefore, this term is 
negligible, and $g$ thus satisfies the Schr\"odinger equation in a harmonic 
potential,
\begin{eqnarray}
 - \frac {\hbar^2 {\ve \nabla}_{\perp}^2} {2m} g + V g=\mu g \ ,
\label{equg1}
\end{eqnarray}
which means that $|g| \propto \exp{[-(x^2+y^2)/2a^2_{\perp}]}$.
From Eq.\,(\ref{mu}), we find that $\mu(\sigma) = \hbar \omega_{\perp}
(1+ 2 a_{\rm sc} \sigma)$. The first term in $\mu$ is the ground-state 
energy of the harmonic oscillator and the second is the interaction energy. 
Thus, $\epsilon(\sigma) = \hbar \omega_{\perp} \sigma (1 +a_{\rm sc}  
\sigma)$ in the low-density limit, and Eq.\,(\ref{poles}) takes the form
\begin{eqnarray}
     \frac {\hbar^2} {2m}
   \left( \frac {\partial \sigma^{1/2}} {\partial z} \right)^2 =
  \left( \frac {\sigma} {2A} U_0 - m u^2 \right) 
 \frac {(\sigma - \sigma_0)^2} {2\sigma} \ ,
\label{poles3d}
\end{eqnarray}
by analogy with Eq.\,(\ref{poles1d}). The results for the profile
and the width of the cloud are analogous to the one-dimensional
case, $u^2 = \sigma_{\rm{min}} U_0/(2mA)$, where $\sigma_{\rm min}$ is
the minimum of $\sigma$ associated with the solitary wave and
\begin{eqnarray}
   \sigma(z) \approx \sigma_{\rm{min}} + (\sigma_0 - \sigma_{\rm{min}})
  \tanh^2 (z/\zeta) \ .
\label{profile2}
\end{eqnarray}
For this problem,
\begin{eqnarray}
   \zeta=2\xi(n_0)[1-(\sigma_{\rm min}/\sigma_0)]^{-1/2}
\label{zeta}
\end{eqnarray}
or $\zeta=2^{1/2}\xi(n_0/2)[1-(\sigma_{\rm min}/\sigma_0)]^{-1/2}$
with $n_0 = \sigma_0/A$.  From Eq. (\ref{poles3d}), we see that the density 
which determines $u^2$ is $\sigma_{\rm{min}}/2A$ and the one that 
determines $\zeta$ is $\sigma_0/2A$. The factor of $1/2$ in these results 
compared with the analogous results for the homogeneous case is due to 
the average of the equilibrium density of atoms across the trap, $n(x,y)$, 
given by $\int n^2(x,y) \, dx dy/ \int n(x,y) \, dx dy$.  Its origin is 
thus completely different from that of the factor of $1/2$ that 
occurs in the Thomas-Fermi limit \cite{KP}. The latter is $\int n(x,y) \, 
dx dy/[n(0,0) \int dx dy]$. It is amusing that the sound speed expressed in 
terms of the density of particles on the axis of the trap is given by 
precisely the same expression, $c=[n(0,0) U_0/(2m)]^{1/2}$, in both the
high- and low-density limits for harmonic transverse trapping potentials.

We argued above that the motion would be quasi-one-dimensional if the
time scale for adjustments in the transverse structure of the cloud is short
compared with the time for passage of the pulse, and we now check the 
consistency of this assumption. In the low-density regime, $\sigma \ll 
\sigma_c$, the characteristic time for adjustment of the profile is on the
order of $\omega_{\perp}^{-1}$, while the time scale for passage of the pulse
is at least of order $\xi/c$. The ratio of these scales is $n U_0/ \hbar 
\omega_{\perp}$, which is much less than unity in this regime, and therefore 
our assumption is consistent.

Another case that can be solved analytically is that of small-amplitude
solitary waves for arbitrary densities.
We expand $\epsilon(\sigma)$ to order $(\sigma - \sigma_0)^3$ 
and write Eq.\,(\ref{poles}) as
\begin{eqnarray}
     \frac {\hbar^2} {2m}
   \left( \frac {\partial \sigma^{1/2}} {\partial z} \right)^2 &\approx&
   \frac 1 2 \epsilon^{''}(\sigma_0)
 (\sigma - \sigma_0)^2 + \nonumber \\
 + \frac 1 6 \epsilon^{'''}(\sigma_0)
 (\sigma &-& \sigma_0)^3 - 
m u^2 \frac {(\sigma - \sigma_0)^2} {2\sigma} \ ,
\label{final}
\end{eqnarray}
where the prime denotes differentiation with respect to $\sigma$.  Clearly, 
two roots of the right side of the above equation are equal to $\sigma_0$. 
The remaining root gives an expression for the velocity $u$ as a function 
of $\sigma_0$ and $\sigma_{\rm min}$, 
\begin{eqnarray}
  u^2 \approx \frac {\sigma_{\rm min}} {m} \left(
   \epsilon^{''}(\sigma_0)
  + \frac { \epsilon^{'''}(\sigma_0)} 3(\sigma_{\rm {min}}-\sigma_0)\right) \ .
\label{root}
\end{eqnarray}
Because $\epsilon^{''}(\sigma)$ depends on $\sigma$,
the velocity of propagation of the wave is not generally equal to the 
bulk velocity at the minimum density, except in the low-density limit, when 
$\epsilon^{''}(\sigma)$ is independent of $\sigma$.  Integrating 
Eq.\,(\ref{final}), we 
can determine the profile of the solitary wave, $\sigma(z)$.  For small 
values of $\sigma - \sigma_0$, Eq.\,(\ref{final}) takes the simple form
\begin{eqnarray}
   \frac {\partial \sigma^{1/2}} {\partial z}  \approx
  \frac  {(\sigma - \sigma_{\rm{min}})^{1/2} } l
 \left( 1 - \frac {\sigma} {\sigma_0} \right)\ ,
\label{profile}
\end{eqnarray}
where the length $l$ is given by $l^2 = \hbar^2/(m \sigma_0
[\epsilon^{''}(\sigma_0) + \sigma_0 \epsilon^{'''}(\sigma_0)/3])$.
Again assuming that $\sigma$ is close to $\sigma_0$, we find that
\begin{eqnarray}
   \sigma(z) \approx \sigma_{\rm{min}} + (\sigma_0 - \sigma_{\rm{min}}) 
  \tanh^2 (z/\zeta)
\label{profile3}
\end{eqnarray}
with $\zeta=l[1-(\sigma_{\rm min}/\sigma_0)]^{-1/2}$. 

We now consider the calculation of $\epsilon(\sigma)$ in the high-density 
limit where the kinetic-energy term is negligible \cite{bp}.  We also 
neglect the last term on the left of Eq.\,(\ref{equilequf}).  The consistency 
of this assumption will be checked below.  With these approximations, we 
write the Thomas-Fermi equation for $g$,
\begin{eqnarray}
     Vg + U_0 |fg|^2 g = \mu g \ .
\label{equg2}
\end{eqnarray}
Thus, $|g|^2$ is a parabola, $|g|^2 \propto [1-(x^2+y^2)/R_{\perp}^2]$.
This form for $|g|$ implies that the potential energy is equal
to $m \omega_{\perp}^2 R_{\perp}^2/6$ and that the interaction energy is
equal to $4 U_0 \sigma/(3 \pi R_{\perp}^2)$.  From Eq.\,(\ref{mu}) we see 
that $\mu(\sigma)$ is the sum of these two terms in this limit. To find the 
explicit dependence of $\mu$ on $\sigma$, we calculate $R_{\perp}$ \cite{KP} 
now.  The density of atoms, $n(x,y,z)$, has the functional form of $|g|^2$,
\begin{eqnarray}
    n(x,y,z)=n(0,0,z) \left(1-\frac{x^2+y^2}{R_{\perp}^2}  \right) \ ,
\label{rho}
\end{eqnarray}
where $n(0,0,z)$ is the density on the axis of the trap.  Thus, the number 
of particles per unit length is given by
\begin{eqnarray}
     \sigma(z)= \int n(x,y,z) \, dx dy = \frac{1}{2} n(0,0,z) \pi 
                R_{\perp}^2 \ .
\label{sigmaz}
\end{eqnarray}
The density on the axis of the trap is given by
\begin{eqnarray}
    n(0,0,z) U_0=\frac{1}{2} m \omega_{\perp}^2 R_{\perp}^2 \ .
\label{condition}
\end{eqnarray}
Thus, we find that $R_{\perp}^2 = 4 a_{\perp}^2 (\sigma a_{\rm sc})^{1/2}$.  
Equation (\ref{mu}) then implies that $\mu(\sigma) = 2 \hbar \omega_{\perp} 
(\sigma a_{\rm sc})^{1/2}$ and therefore $\epsilon(\sigma) = (4/3) 
\hbar \omega_{\perp} (a_{\rm sc} \sigma^3)^{1/2}$. Going back to the width, 
$\zeta$, of the solitary waves, we find that $\zeta = (12/5)^{1/2}\xi(n_0) 
[1-(\sigma_{\rm min}/ \sigma_0)]^{-1/2}$ where $n_0 = \sigma_0/ 
(\pi R_{\perp}^2)$.

In considering the limits of validity of our calculation in the high-density 
regime, we observe that the time scale for adjustment of the profile of the 
pulse is $\sim\,R_{\perp}/c$ and that the time for passage of the pulse 
is $\sim\,\zeta/c$.  If the motion is to be essentially one dimensional, 
$R_{\perp}$ must be much smaller than $\zeta$.  In the Thomas-Fermi 
approximation, $R_{\perp}$ is larger than $a_{\perp}$ due to the repulsive 
interactions between the atoms.  This can be seen from the formula for the 
radius $R_{\perp} = 2 a_{\perp} (\sigma a_{\rm sc})^{1/4}$ derived above.  
To satisfy the condition $\zeta \gg R_{\perp}$, the quantity $\sigma 
a_{\rm sc}$ must be large compared with $(3/10)^2 (\sigma_0 a_{\rm sc})^{-1} 
[1-(\sigma_{\rm min}/ \sigma_0)]^{-2}$.  Thus, if $\delta \sigma = \sigma_0 
- \sigma_{\rm min}$ is the amplitude of the disturbance, we see that 
$\delta \sigma/ \sigma_0 \ll 3 /(10 \sigma_0 a_{\rm sc})$.  This indicates 
that the amplitude of the solitary wave, $\delta \sigma$, must be extremely 
small if the one-dimensional approximation is to be valid when $\sigma 
\gg \sigma_c$, and therefore our small-amplitude treatment given above
is applicable.  For the experimental conditions of Ref.\,\cite{andrews},
the number of particles in the trap is $N \sim 5 \times 10^6$, 
and the length of the trap in the axial direction $L$ is $\sim 450$ $\mu$m, 
implying that $\sigma_0 \sim 10^8$ cm$^{-1}$.  Validity of the one-dimensional
approximation therefore requires that $\delta \sigma/ \sigma_0 \ll 0.01$.  
Finally, we check the consistency of the assumption made in deriving 
the Thomas-Fermi equation for $g$.  The last term on the left of 
Eq.\,(\ref{equilequf}) has an upper bound of $\hbar^2/(2 m \zeta^2)$.  This 
term is approximately equal to $\hbar^2/(2 m \xi^2) (\delta \sigma/
\sigma_0) \ll \hbar^2/(2 m \xi^2) \sim n U_0$.  Therefore, this term is
indeed negligible.
 
Our calculations indicate that the most favorable conditions for observing 
solitary waves in trapped Bose-condensed gases occur when the density is 
sufficiently low to permit a perturbative treatment of particle interactions. 
In this case, one-dimensional behavior persists even for large-amplitude 
solitary waves.  Low-density systems have the further advantage that the 
coherence length is correspondingly large.  Since the coherence length 
determines the size of solitary waves, this simplifies resolution of these 
structures.  However, a lower density of particles makes the detection of 
these effects more difficult because of the lower signal.

We now wish to estimate the experimental conditions required for observation 
of solitary waves in the low-density regime. The spatial resolution of 
current experiments using direct imaging methods is $\sim 4$ $\mu$m 
\cite{Wolfgang}. As a theoretical estimate for the width of the solitary 
wave, we take the full width of the dip in the density profile at half
the maximum depression, that is the distance between points where
$\sigma=(\sigma_0+\sigma_{\rm min})/2$. This is given by $2 \zeta 
\tanh^{-1}(1/\sqrt{2}) \approx 3.5 \, \xi(n_0) [1-(\sigma_{\rm min}/
\sigma_0)]^{-1/2}$, which must exceed the experimental resolution if 
solitary waves are to be observable. This leads to the condition 
\begin{eqnarray}
   \xi(n_0) {\,\raisebox{-.5ex}{$\stackrel{>}{\sim}$}\,} 
  [1-(\sigma_{\rm min}/\sigma_0)]^{1/2} \mu{\rm m},
\label{xi0}
\end{eqnarray} 
or $\xi(n_0) {\, \raisebox{-.5ex}{$\stackrel{>}{\sim}$}\,} 1$ $\mu$m for
solitary waves with $\sigma_{\rm min}/\sigma_0 \ll 1$. For the coherence
length to exceed 1 $\mu$m, the density per unit volume must be less than
$n_{\rm obs} \approx 10^{13}$ cm$^{-3}$. The corresponding number of particles
per unit length $\sigma$ is $n_{\rm obs} A \approx n_{\rm obs} \pi 
a_{\perp}^2$. For a transverse trapping frequency of 240 Hz, 
which is a typical value of the MIT trap \cite{andrews,hans}, 
the oscillator length, $a_{\perp}$, in this direction is $\sim\,1.4$ 
$\mu$m, implying that the number of particles per unit length
must be less than $8 \times 10^5$ cm$^{-1}$ for structures to be
observable.  This value corresponds to $\sigma_c/5$, and therefore the
coupling would be weak. Let us now calculate the number of particles
which should be used in the MIT trap for $\sigma$ to have the value
$8 \times 10^5$ cm$^{-1}$. The kinetic energy along the axial direction
is typically of order $\hbar^2/2 m L^2$, and the interaction energy is
$\sim\,\sigma U_0/(\pi a_{\perp}^2)$. The ratio of these quantities is
$\sim\,(a_{\perp}/L)^2 (8\sigma a_{\rm sc})^{-1}$. Thus for $L \gg a_{\perp} 
/(8\sigma a_{\rm sc})^{1/2}$, the kinetic energy along the $z$ direction 
is negligible, and the Thomas-Fermi approximation \cite{bp} may be used
to determine the structure along the axis of the trap.  
We shall assume this to be the case and subsequently will check the 
consistency of this assumption.  Under these conditions, the Thomas-Fermi 
approximation may be used to calculate the $z$-dependence of $\sigma$, even 
though it may not be used to calculate the structure of the cloud in the 
transverse directions.  In the presence of a confining potential in the 
$z$ direction, the Thomas-Fermi condition is that the sum of the chemical 
potential and the $z$-dependent part of the trapping potential should be 
constant.  Since the chemical potential is $\mu = \hbar \omega_{\perp} 
(1 + 2 \sigma a_{\rm sc})$ in the low-density limit, this condition is
\begin{eqnarray}
    \hbar \omega_{\perp} [1 + 2 \sigma(z) a_{\rm sc}]
 + \frac 1 2 m \omega_z^2 z^2 ={\rm constant},
\label{tf1d}
\end{eqnarray}
where $\omega_z$ is the frequency of the trapping potential along the
$z$ axis.  We equate the value of the left side of Eq.\,(\ref{tf1d}) at 
$z=0$ to the value at $z=Z$, where $Z$ is the distance from the center 
of the cloud to its edge, $Z=L/2$.  Since $\sigma$ vanishes at the edges 
of the cloud, we get
\begin{eqnarray}
  2  \hbar \omega_{\perp} \sigma(z=0) a_{\rm sc} = \frac 1 2 m \omega_z^2 Z^2,
\label{TF}
\end{eqnarray}
where $\sigma(z=0)$ is the number of particles per unit
length at the center of the cloud. Solving the above equation for $Z$, 
we find
\begin{eqnarray}
   Z &=& 2 \frac {a_z^2} {a_{\perp}} [\sigma(z=0) a_{\rm sc}]^{1/2}
\nonumber \\ &=&
  2 \sqrt{\pi} (n_{\rm obs} a_{\rm sc})^{1/2} a_z^2 \ ,
\label{ZZ}
\end{eqnarray}
where $a_z = (\hbar/m \omega_z)^{1/2}$ is the oscillator length in the
axial direction. From Eqs.\,(\ref{tf1d}) and (\ref{ZZ}) we see that
the number of particles per unit length can be written as
\begin{eqnarray}
  \sigma(z) = \sigma(z=0) \left(1 - \frac {z^2} {Z^2} \right)\ .
\label{sigmaa}
\end{eqnarray}
The total number of particles is $N = \int_{-Z}^{Z} \sigma(z) dz$; 
thus from Eqs.\,(\ref{ZZ}) and (\ref{sigmaa}) we find 
\begin{eqnarray}
  N = \frac 4 3  \sigma(z=0) Z = 
 \frac {8 \pi^{3/2}} 3 n_{\rm obs}^{3/2} (a_z a_{\perp})^2 a_{\rm sc}^{1/2} \ .
\label{number}
\end{eqnarray}
For the MIT trap, the frequency in the axial direction is $\sim 7.9$ Hz 
\cite{hans} and therefore $a_z \sim\,7.5$ $\mu$m; the 
total number of particles, $N$, should be 
equal to $\sim\,2 \times 10^3$ if $\sigma(z=0)$ is to be $8 \times 10^5$ 
cm$^{-1}$. The total width of the trap in the $z$ direction would then be 
$L \approx 38$ $\mu$m.  Returning to our assumption, we see that for $\sigma 
= \sigma_c/5$ (the highest value of $\sigma$ for which these structures are 
observable), the condition $L \gg a_{\perp}/(8 \sigma a_{\rm sc})^{1/2}$ 
implies that $L$ must be much larger than $0.8a_{\perp}$. This is indeed true
and the Thomas-Fermi approximation is valid along the $z$ axis of the cloud.

The development of traps with stronger transverse confinement, such as 
the optical dipole trap of the MIT group \cite{new}, promises to 
facilitate experiments on solitary waves since, for a given value of 
the dimensionless coupling $\sigma a_{\rm sc}$, the corresponding densities 
will be higher.  However, the disadvantage is that the higher density will 
result in shorter length scales for structures. 
 
In this paper we have discussed the propagation of solitary waves in elongated 
traps and have estimated characteristic sizes of these structures. Clearly, 
many questions remain.  These include the stability of these structures, 
the role they play in other phenomena (such as dissipation), and the 
question of how pulses propagate when the motion is not 
quasi-one-dimensional.
\\ \\
   Helpful discussions with H. Fogedby, W. Ketterle, H.-J. Miesner,
M. Saffman and H. Smith
are gratefully acknowledged. G.M.K. would like to thank the 
Foundation of Research and Technology, Hellas (FORTH) for its hospitality.

\end{document}